\documentclass[reprint,prd,superscriptaddress,nofootinbib,amsmath,amssymb]{revtex4-2}
\usepackage{xparse, physics, graphicx, dcolumn, bm, amsmath, amssymb, mathtools, cancel, dsfont, enumitem, braket, tikz, color}
\usepackage[normalem]{ulem}
\usepackage[linktocpage=true]{hyperref}
\usepackage[all]{hypcap}
\usepackage{cleveref}
\usepackage{siunitx}
\usepackage{verbatim}

\DeclareSIUnit{\Mpc}{Mpc}
\newcommand{\MPl}{M_\text{Pl}}
\newcommand{\msoft}{m_{\text{soft}}}
\newcommand{\hc}{\textup{h.c.}}
\newcommand{\ii}{\mathrm{i}}
\NewDocumentCommand{\so}{m!O{}}{\mathfrak{so}(#1)_{\text{#2}}}
\NewDocumentCommand{\su}{m!O{}}{\mathfrak{su}(#1)_{\text{#2}}}
\NewDocumentCommand{\SO}{m!O{}}{\textup{SO}(#1)_{\text{#2}}}
\NewDocumentCommand{\SU}{m!O{}}{\textup{SU}(#1)_{\text{#2}}}
\NewDocumentCommand{\U}{m!O{}}{\textup{U}(#1)_{\text{#2}}}
\NewDocumentCommand{\Z}{O{}}{\mathbb{Z}_{#1}}
\newcommand{\Lag}[1][]{\mathcal{L}_\text{#1}}
\newcommand{\defeq}{\equiv}

\newcommand{\cmt}[1]{}

\newcommand{\sref}{\sigma_*}
\newcommand{\tref}{t_*}
\newcommand{\kref}{k_*}

\newcommand{\mVac}{m_{\chi, 0}}
\newcommand{\mInf}{m_{\chi, I}}
\newcommand{\OmegaGWaniso}{\Omega_\text{GW, aniso}}
\newcommand{\sr}{\mathrm{sr}}

\let\deltafunc\delta
\DeclareDocumentCommand\delta{}{\trigbraces{\deltafunc}}
\let\thetafunc\Theta
\DeclareDocumentCommand\Theta{}{\trigbraces{\thetafunc}}

\makeatletter
\let\l@subsubsection\@gobbletwo
\makeatother

\begin{document}
\title{
Spectator Composes a Gravitational Canon:\\
Spectator-field-triggered Phase Transition During Inflation\\ and its Anisotropic Gravitational Wave Signals
}

\author{Yunjia Bao}
\email{yunjia.bao@uchicago.edu}
\affiliation{Department of Physics, Enrico Fermi Institute, Leinweber Institute for Theoretical Physics, Kavli Institute for Cosmological Physics, University of Chicago, Chicago, IL 60637, USA}

\author{Keisuke Harigaya}
\email{kharigaya@uchicago.edu}
\affiliation{Department of Physics, Enrico Fermi Institute, Leinweber Institute for Theoretical Physics, Kavli Institute for Cosmological Physics, University of Chicago, Chicago, IL 60637, USA}
\affiliation{Kavli Institute for the Physics and Mathematics of the Universe (WPI), The University of Tokyo Institutes for Advanced Study, The University of Tokyo, Kashiwa, Chiba 277-8583, Japan}

\begin{abstract}
We propose a general framework in which a phase transition is triggered during cosmic inflation by the slow-roll dynamics of a spectator field. The topological defects formed at the transition are inflated outside the horizon, reenter it after inflation, and can subsequently generate characteristic gravitational-wave (GW) signals. Quantum fluctuations of the spectator field modulate the timing of the transition, imprinting large-scale anisotropies in the resulting GW background. As an explicit realization, the spectator field may be identified with the Higgs field in a supersymmetric Standard Model. More generally, our framework applies to a wide class of spectator-modulated phenomena, providing a generic mechanism for producing anisotropic GW signals.
\end{abstract}

\maketitle

\tableofcontents

\section{Introduction}
Symmetry and its spontaneous breaking are ubiquitous in particle physics. A prominent example is the electroweak symmetry, which is spontaneously broken down to the electromagnetic symmetry. In grand unified theories, gauge symmetries such as $\SU{5}$ and $\SO{10}$ are broken to the Standard Model (SM) gauge symmetry~\cite{Georgi:1974sy, Fritzsch:1974nn}. In the Peccei–Quinn solution to the strong CP problem~\cite{Peccei:1977hh, Peccei:1977ur}, a global $\U{1}$ symmetry is spontaneously broken.

Symmetry breaking is often accompanied by the formation of topological defects, including magnetic monopoles, cosmic strings, and domain walls. If the symmetry breaking occurs before the observable inflation, these defects are inflated away and leave no observable imprint. Conversely, if the symmetry is broken after inflation, topological defects may give rise to observable signals, such as strong ionization by magnetic monopoles, anisotropies in the cosmic microwave background induced by cosmic strings, or gravitational waves (GWs). See Refs.~\cite{Vilenkin:1984ib, Vilenkin:2000jqa} for reviews.

Less explored, but potentially more striking, is the possibility that symmetry breaking occurs \textit{during} the observable inflation. In this case, the resulting topological defects are inflated outside the horizon, rendering them effectively frozen by causality. After inflation ends, these defects reenter the horizon and subsequently evolve.\footnote{
A similar evolution of topological defects may also arise if the phase transition occurs between two stages of inflation, where the latter stage could be a period of thermal inflation~\cite{Yamamoto:1985rd, Lyth:1995ka}.
}
Observable consequences of this scenario include an enhanced abundance of axions~\cite{Baratella:2018pxi, Redi:2022llj, Harigaya:2022pjd} and GWs with characteristic spectral features~\cite{Martin:1996ea, Martin:1996cp, Lazarides:2021uxv, Maji:2022jzu, Bao:2024bws, Maji:2024cwv}.

In the literature, phase transitions during inflation have often been realized by coupling the symmetry-breaking sector directly to the inflaton field~\cite{Senoguz:2015lba, An:2020fff, Chakrabortty:2020otp, Lazarides:2021uxv, An:2022cce, Maji:2022jzu, An:2023idh, An:2023jxf, Maji:2024cwv, An:2024oui}. Such scenarios typically require a large excursion of the inflaton field value in order for the mass of the symmetry-breaking field to vary significantly during inflation.

In this work, we point out an alternative possibility in which a phase transition is triggered by the dynamics of a field whose energy density is subdominant during inflation, \textit{i.e.}, a spectator field. In this setup, the timing of the phase transition is modulated by the inflationary quantum fluctuations of the spectator field. As a result, the horizon reentry of the resulting topological defects, and hence any signals they produce, is spatially modulated. In particular, if the topological defects emit GWs, this mechanism naturally generates large-scale anisotropies in the GW background.

Anisotropies in GWs sourced by scalar perturbations~\cite{Bartolo:2019oiq, Bartolo:2019yeu, Dimastrogiovanni:2022eir, Chen:2022qec, Malhotra:2022ply, Li:2023qua, Li:2023xtl, Wang:2023ost, Yu:2023jrs, Ruiz:2024weh, Rey:2024giu, Li:2025met, Yu:2025jgx, Bodas:2025wef} and by phase transitions~\cite{Geller:2018mwu, Kumar:2021ffi, Bodas:2022zca, Bodas:2022urf} have been studied previously. Compared to existing models, one of our realizations based on cosmic strings allows for the largest possible GW anisotropy while remaining consistent with constraints from (iso)curvature perturbations and their non-Gaussianity.

This paper is organized as follows. In \cref{sec:PTduringInflation}, we present the general mechanism by which a phase transition during inflation is triggered by a light spectator field whose spectrum is discussed in \cref{subsec:lightSpectator}. Fluctuations of the spectator field value leads to a fluctuation in the phase transition timing during inflation, as discussed in \cref{subsec:PTduringInflation}. We then study the GW signal of cosmic strings in \cref{sec:GWfromString} and that of domain walls in \cref{sec:GWfromWall}. Their respective GW anisotropy signals are discussed in \cref{subsec:StringSignal,subsec:WallSignal}, and their constraints are discussed in \cref{subsec:StringConstraints,subsec:WallConstraints}. In \cref{app:two fields}, we point out that a two-field symmetry breaking allows the defect-forming field during inflation to differ from that in the vacuum. This is then realized in a supersymmetric embedding of the mechanism in which the SM Higgs field functions as the spectator field, as discussed in \cref{app:SUSY}. We conclude in \cref{sec:conclusion}.

\begin{figure*}
    \centering
    \includegraphics[width=\linewidth]{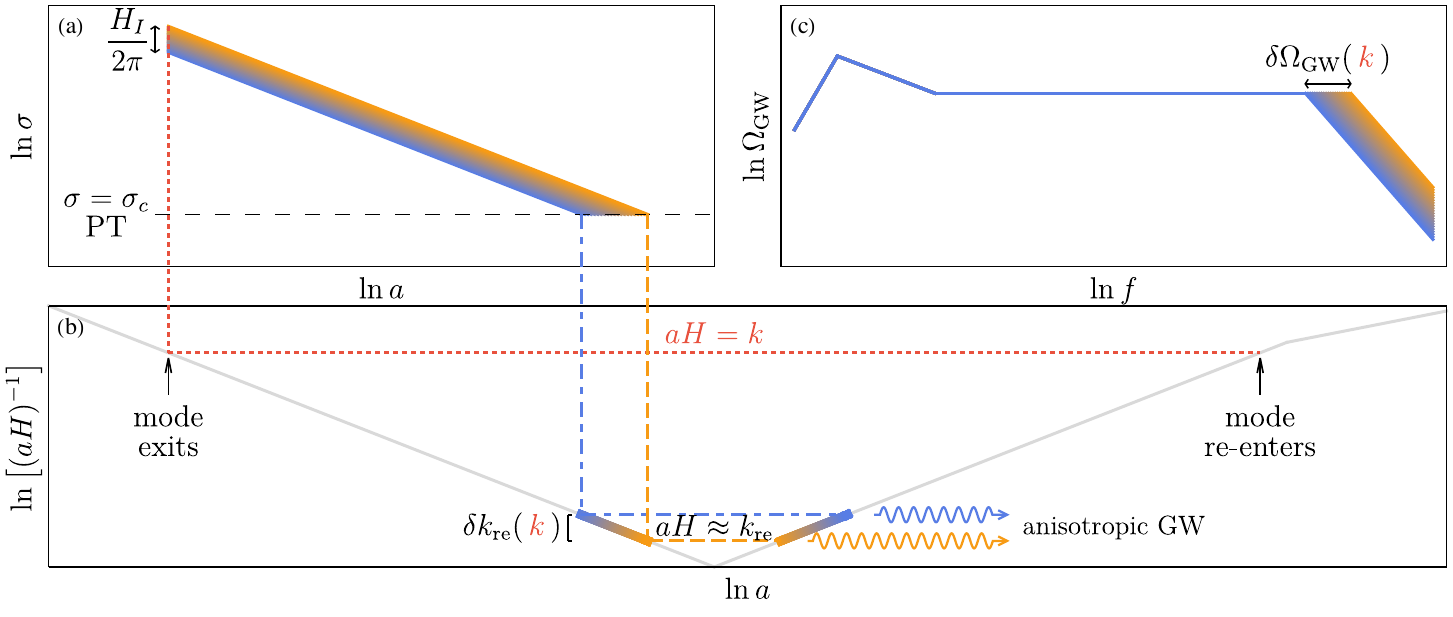}
    \caption{
    Schematic of the mechanism. 
    \textbf{Top Left (a)}: A fluctuation of the spectator field value $\sigma$ early in the inflation epoch imprints as a fluctuation in the phase transition (PT) timing during inflation. This changes when PT happens in different Hubble patches.
    \textbf{Bottom (b)}: The fluctuation in PT timing at large distance $k^{-1}$ changes the horizon size $k_\text{re}^{-1}$ when the PT defects reenter. When the mode $k$ reenters the horizon, a large-scale anisotropic gravitational-wave (GW) signal appears if the GW spectrum $\Omega_\text{GW}$ depends on $k_\text{re}$.
    \textbf{Top Right (c)}: GW spectrum is modulated because of the reentry timing. Here, we take GWs produced by cosmic strings as an example. Modulation in $k_\text{re}$ changes the UV rolloff scale. This modulation is at large scales because it comes from the early fluctuation of $\sigma$ during inflation. 
    }
    \label{fig:schematic}
\end{figure*}

\section{Phase Transition During Inflation Triggered by Light Spectator}\label{sec:PTduringInflation}

Physics beyond the SM typically contains new scalar fields. A notable example is supersymmetric theories, where the scalar partners of the SM fermions are predicted. In this section, we discuss how those fields can trigger phase transition during inflation.

\subsection{Light Spectator During Inflation}
\label{subsec:lightSpectator}

If the inflationary Hubble scale $H_I$ is larger than their masses, they may be displaced from the minimum and obtain quantum fluctuations.

We consider the case where a spectator field $\sigma$ has a mass squared $m_\sigma^2$ of $\order{0.01-0.1} H_I^2$ during inflation. The mass may be a vacuum mass or a so-called Hubble-induced mass that is given by the coupling of $\sigma$ with the Ricci scalar or the inflaton potential. Such a field may undergo a slow roll while obtaining de Sitter quantum fluctuations. Given an $m_\sigma^2 \sigma^2/2$ potential, the equation of motion for $\sigma$ is%
\footnote{
Instead of a quadratic mass, we may consider a non-harmonic potential with a negligible quadratic term. In this case, the attractor behavior can drive $\sigma$ to the field value where a slow roll occurs~\cite{Dimopoulos:2003ss, Harigaya:2012up, Kalia:2025uxg}.
}
\begin{equation}
    \begin{multlined}
        \ddot{\sigma} + 3H_I \dot{\sigma} \approx - m_\sigma^2 \sigma \\
        \implies \sigma(t) \approx \sref \exp(-\beta H_I \Delta t), \quad 
        \beta \defeq \frac{3}{2} - \sqrt{\frac{9}{4} - \frac{m_\sigma^2}{H_I^2}},
    \end{multlined}
\end{equation}
in which $\sref$ denotes the field value at some reference time $\tref$ during inflation and $\Delta t$ is the time past from $\tref$. Because the scale factor $a \propto e^{H_I t}$ during inflation, the classical rolling of $\sigma$ scales as $\sigma \propto a^{-\beta}$.
Taking the comoving scale that exits the horizon at each time, $k = a H_I$, as a time variable,
\begin{equation}
    \sigma(k) = \sref \qty(\frac{\kref}{k})^{\beta},
    \label{eqn:sigmaOfK}
\end{equation}
with some reference scale $\kref$ such that $\sigma(\kref) = \sref$.

On the other hand, light fields during inflation inherit fluctuations of order $\delta\sigma \sim H_I/(2\pi)$ when each mode exits the horizon~\cite{Mukhanov:1981xt, Hawking:1982cz, Starobinsky:1982ee, Guth:1982ec, Bardeen:1983qw}. Outside the horizon, $\delta \sigma$ and $\sigma$ scales in time in the same way.
Thus, the primordial dimensionless spectrum of $\sigma$ is blue-tilted as 
\begin{equation}
    \Delta_\sigma^2(k) \approx \qty(\frac{\delta \sigma}{\sigma(k)})^2 = \qty(\frac{H_I}{2\pi \sref})^2 \qty(\frac{k}{\kref})^{2\beta}. 
    \label{eqn:Psigma2}
\end{equation}

\subsection{Phase Transition During Inflation}
\label{subsec:PTduringInflation}
We propose the possibility that $\sigma$ triggers a phase transition during inflation. The phase transition may lead to various stochastic GW signals if it 1) produces topological defects, 2) generates large scalar fluctuations, or 3) has a strong enough phase transition strength. The precise GW spectrum depends on the dominant GW source. However, GW anisotropy is a generic feature for this class of models. We will show this with explicit examples. 

We first discuss how a phase transition can be triggered by a spectator during inflation. We consider the following Lagrangian
\begin{equation}
    - \Lag \supset \frac{1}{2} \qty(\lambda_{\sigma\chi}^2 \sigma^2 - m_\chi^2) \chi^2,
\end{equation}
in which the symmetry-breaking field $\chi$ has a tachyonic mass $m_\chi^2$ and couples quadratically to $\sigma$. Here, $m_\chi^2$ has two potential contributions: 1) a vacuum mass $\mVac^2$ and 2) a Hubble-induced mass $ \sim c_\chi H_I^2$ with $c_\chi \sim \order{1}$ to ensure that classical rolling of $\chi$ dominates during inflation. For high-scale inflation, $c_\chi H_I^2 \gg \mVac^2$ will affect the inflationary dynamics of $\chi$. On the other hand, the vacuum expectation value (VEV) of $\chi$ today is controlled by $\mVac$. We keep track of the two masses with the notation $\mVac^2$ and $\mInf^2 \equiv \mVac^2+ c_\chi H_I^2$ in what follows.

Initially, when $\sigma$ has a large field value, $\chi$ has a positive mass and a vanishing field value. As $\sigma$ slowly rolls to the origin and passes the critical value $\sigma_c = \mInf/\lambda_{\sigma\chi}$, $\chi$ obtains a non-zero field value and undergoes a phase transition. When this happens at the comoving Hubble scale $k_\text{re}$, one expects some spectral feature at this scale as it reenters the horizon after inflation.

The reentry timing of the phase transition during inflation depends on 1) the initial value of $\sigma$ and 2) the number of $e$-foldings from the phase transition to the end of inflation. These two freedoms allow us to treat $\sref$ and $\kref$ in \cref{eqn:Psigma2} as free parameters. For convenience, we may pick $\kref = k_\text{re}$ and $\sref = \sigma_c$. From \cref{eqn:sigmaOfK}, we find that the fluctuation in $k_\text{re}$ is related to the fluctuation in $\sigma$ at comoving scale $k$,
\begin{equation}
    \frac{\delta k_\text{re}}{k_\text{re}} (k) = \frac{1}{\beta} \Delta_\sigma(k) = \frac{1}{\beta} \qty(\frac{H_I}{2\pi \sigma_c}) \qty(\frac{k}{k_\text{re}})^\beta.
    \label{eqn:deltaKRe}
\end{equation}
Therefore, the long-wavelength fluctuation of $\sigma$ imprints itself as long-wavelength modulation of $k_\text{re}$. If the GW signal from phase transition depends on $k_\text{re}$, then a large-scale anisotropic GW is expected as sketched in \cref{fig:schematic}. When $m_\chi$ is dominated by the Hubble-induced mass, $H_I/(2\pi \sigma_c) = \lambda_{\sigma \chi} / (2\pi c_\chi^{1/2})$, and the anisotropy is naturally large. We consider the parameter region with $\delta k_\text{re}/k_\text{re} <1$. Otherwise, the dynamics of $\sigma$ during inflation is dominated by quantum fluctuations rather than the classical rolling, and our computation is not applicable.

\section{Gravitational-Wave Anisotropy Signal from Cosmic Strings}\label{sec:GWfromString}

\begin{figure*}
    \centering
    \includegraphics[width=\linewidth]{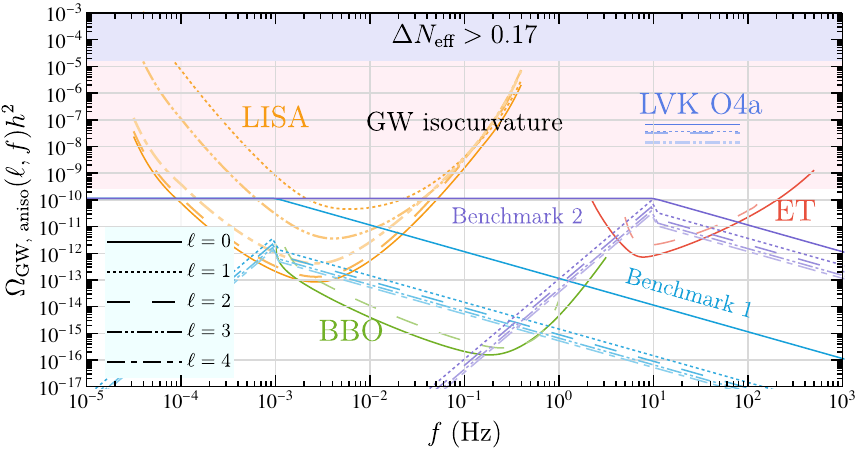}
    \caption{
    Benchmark spectra for gravitational-wave (GW) anisotropy produced by spectator-modulated cosmic-string reentry. While the string tension $\mu$ controls the plateau of the isotropic signal, the comoving reentry scale $k_\text{re}$ of the string alters the UV rolloff scale. A modulation in $k_\text{re}$ leads to an anisotropic signal $\OmegaGWaniso(\ell,f)$ in higher multipoles on top of the isotropic background. Our Benchmark 1 is taken at $(\sqrt{\mu}, k_\text{re}/(2\pi), \beta, H_I/(2\pi \sigma_c)) = (\SI{3E14}{\GeV}, \SI{E-3}{\Hz}, 10^{-1}, 10^{-1})$, and Benchmark 2 is taken at $(\sqrt{\mu}, k_\text{re}/(2\pi), \beta, H_I/(2\pi \sigma_c)) = (\SI{3E14}{\GeV}, \SI{10}{\Hz}, 10^{-2}, 10^{-2})$. The current exclusions from LIGO-Virgo-KAGRA's first part of the fourth observing run (LVK O4a) \cite{LIGOScientific:2025bkz}, CMB measurement of $\Delta N_\text{eff}$ \cite{Planck:2018vyg}, GW isocurvature perturbations~\cite{Planck:2018jri}, and the prospects of future searches (LISA \cite{LISACosmologyWorkingGroup:2022kbp}, BBO \cite{Cui:2023dlo, Schmitz:2020syl}, and ET \cite{Cui:2023dlo, Mentasti:2020yyd, Schmitz:2020syl}) of GW anisotropy are provided for comparison. Future detector reaches are power-law-integrated sensitivity curves with a signal-to-noise threshold of $1$ and an integration time of $1$ year. Details of the GW constraints and prospects are provided in \cref{app:GWSensitivity}.}
    \label{fig:benchmarks}
\end{figure*}

Large-scale fluctuations of $\sigma$ may produce GW anisotropies. The multipole index $\ell$ of the anisotropic signal roughly scales as $\ell \approx k D_*$, in which $D_*$ denotes the comoving distance to the emission surface of GWs. As $D_*$ in radiation domination (RD) remains almost unchanged for earlier time, the emission surface of GWs almost coincides with the last scattering surface of the cosmic microwave background (CMB), $D_* \approx \SI{1.4E4}{\Mpc}$. We will be interested in low-multipole ($\ell \lesssim 5$) anisotropies that can be observed by future GW observatories.

\subsection{GW Signal}
\label{subsec:StringSignal}
Consider a scalar field $\chi$ charged under some $\U{1}$ gauge symmetry that couples to the spectator field, 
\begin{equation}
    - \Lag \supset \qty(\lambda_{\sigma \chi}^2 \sigma^2 - m_\chi^2) \abs{\chi}^2 + \frac{\lambda_\chi^2}{4} \abs{\chi}^4.
\end{equation}
When the phase transition occurs during inflation,
the Kibble-Zurek mechanism produces cosmic strings. After inflation, these strings will have a tension of $\mu = \pi v_\chi^2$, where the VEV of $\chi$ is $v_\chi = 2\mVac/\lambda_{\chi}$.

The model may be extended so that the symmetry-breaking field during inflation is different from that at the vacuum, which helps identify $\sigma$ with the Higgs field 
in the minimal supersymmetric SM (MSSM); see \cref{app:SUSY}.  

The GW signature of this string network should be similar to that of a typical gauge string network, but the UV rolloff depends on the reentry Hubble scale of the inflated strings instead of the phase transition Hubble scale \cite{Guedes:2018afo, Cui:2019kkd}. For illustrative purposes, we simplify the GW spectrum as a broken power law%
\footnote{Here, we focus on the GW spectrum around string reentry as it generates most of the GW anisotropy. For string loops produced later, the $g_*$-dependent factor should be adjusted and increase as $g_*$ decreases.}
\begin{equation}
    \begin{multlined}
        \Omega_\text{GW} \approx 2\times 10^{-2} \Omega_\text{rad} \qty(\frac{g_{*,0}}{g_{*,\text{re}}})^{1/3} \sqrt{\frac{\mu}{\MPl^2}} \\
        \times 
        \begin{dcases}
            1 + \qty(\frac{2\pi f}{k_\text{re}})^3, & f < \frac{k_\text{re}}{2\pi}, \\
            \qty(\frac{k_\text{re}}{2\pi f}), & f \geq \frac{k_\text{re}}{2\pi}, \\
        \end{dcases}
    \end{multlined}
\end{equation}
in which $\Omega_\text{rad} = 9.14\times10^{-5}$ is the current fractional density of photons and massless neutrinos, $f$ is the frequency of the GWs, $g_{*,0(\text{re})}$ denotes the effective number of relativistic degrees of freedom today (at defects' reentry), and we assume that the $k_\text{re}^{-1}$-sized string loop decays during radiation domination. The $k_\text{re}^{-1}$-sized loops initially produce the $f^3$ IR spectrum before shrinking and later generate the $f^{-1}$ UV spectrum, as the energy decreases in proportion to their radii. The modulation on the GW spectrum due to $\sigma$ fluctuation is
\begin{equation}
   \delta \Omega_\text{GW} = \frac{\delta k_\text{re}}{k_\text{re}} \pdv{\Omega_\text{GW}}{\ln k_\text{re}}.
\end{equation}
The GW anisotropy signal can be obtained using \cref{eqn:deltaKRe},
\begin{align}
    &\OmegaGWaniso(\ell,f) \sim 2\times 10^{-2} \Omega_\text{rad} \sqrt{\frac{\mu}{\MPl^2}} \qty(\frac{g_{*,0}}{g_{*,\text{re}}})^{1/3} \\
    & \times \frac{\lambda_{\sigma\chi} H_I}{2\pi \mInf} \frac{1}{\beta} \qty(\frac{\ell}{k_\text{re} D_*})^\beta \times
    \begin{dcases}
        3 \qty(\frac{2\pi f}{k_\text{re}})^3, & f < \frac{k_\text{re}}{2\pi}, \\
        \frac{k_\text{re}}{2\pi f}, & f > \frac{k_\text{re}}{2\pi}.
    \end{dcases} \nonumber
\end{align}
Hence, one expects a slightly blue-tilted anisotropy signal, a telltale feature of the light spectator field. We offer two benchmark spectra against exclusion and reach from current and future GW observatories in \cref{fig:benchmarks}.

\subsection{Constraints}
\label{subsec:StringConstraints}
The gauge string's energy density around the string reentry is mostly radiated into GWs and can produce dark-radiation isocurvature perturbations. The fluctuation of the energy density around the matter-radiation equality is at most $\Delta_\zeta$ at large scales; thus, the large-scale anisotropy of GW energy density is at most $\lesssim \Delta_\zeta \Omega_\text{rad} h^2 \sim 10^{-10}$ in the current universe. This computation is detailed below where we find
\begin{equation}
    \delta \Omega_\text{GW}(f_\text{peak}) h^2 \lesssim  2.8\times 10^{-10},
\end{equation}
where $f_\text{peak}$ is the GW frequency at which the anisotropy is maximized. For cosmic strings, $f_\text{peak}\simeq k_\text{re}/(2\pi)$.  Because the isocurvature perturbations are nonlinearly related to the nearly Gaussian $\sigma$ fluctuation, this leads to a local-type non-Gaussianity as well. However, our model evades the current \textit{Planck} constraint on such non-Gaussianity~\cite{Planck:2019kim}. 

\subsubsection{Isocurvature Power Spectrum}
Here, we assume that, for gauge strings, the string loops decay entirely into  gravitational waves. This is sensible because the gauge and Higgs bosons of the $U(1)$ symmetry breaking have larger masses than the Hubble scale that determines the typical length scale of string reconnection and loop production. Because string loops redshift like matter rather than radiation, their energy density during RD can contribute significantly to the energy density of the network even when the string tension is small. In fact, the typical decay rate of string loops into GWs is $\Gamma_\text{str}\sim \mu H_p/\MPl^2 \ll H_p $ in which $H_p$ denotes the Hubble scale around string loop production. The typical energy density of the string loops scales as 
\begin{equation}
    \eval{\rho_\text{loop}}_{t = \Gamma_\text{str}} \approx \mu H_p^2 \qty(\frac{a(H_p)}{a(\Gamma_\text{str})})^3 = \frac{\sqrt{\mu} \MPl}{t^2} \approx \rho_\text{GW}.
\end{equation}
Hence, during RD, a string network that reaches the scaling regime shall produce an energy density
\begin{equation}
    \rho_\text{long str} \approx \mu H^2,~~
    \rho_\text{GW} \approx \gamma \sqrt{\mu} \MPl H^2,
\end{equation}
in which $\gamma \approx 2 \times 10^{-2}$ is a dimensionless number obtained from numerical simulations \cite{Quashnock:1990wv, Blanco-Pillado:2017oxo, Sousa:2020sxs}. 
Both the GW density produced during the scaling regime and the long-string energy density are independent of the timing of the horizon reentry of the strings, and therefore do not contribute to cosmic perturbations.
On the other hand, the energy density of GWs produced by string loops right after the horizon reentry depends on the timing of the reentry and contributes to cosmic perturbations.

The curvature perturbations can be computed using the $\delta N$ formalism~\cite{Sasaki:1995aw, Wands:2000dp, Lyth:2004gb}. To relate GW's perturbation to observational constraints, we note that the de Sitter fluctuation of $\sigma$ generates uncorrelated isocurvature fluctuations in the free-streaming GW. Hence, we may recast the neutrino density isocurvature (NDI) constraint from \textit{Planck} \cite{Planck:2018jri}. The NDI perturbation is defined as 
\begin{equation}
    S_\text{NDI} \defeq 3\qty(\zeta_\nu - \zeta_\gamma),
\end{equation}
in which $\zeta_{\nu(\gamma)}$ are the curvature perturbations of neutrinos (photons). Here, we reinterpret $\zeta_\nu$ as the curvature perturbations from any free-streaming species. In particular, if neutrino perturbations are generated from inflaton fluctuations, and GWs are produced from $\sigma$ fluctuations, then the NDI perturbation comes from the GW from cosmic strings' reentry. To compute $\zeta_\nu$, we take both the initial and final slices in the scaling regime after the string reentry. The initial flat slice coincides with the uniform density slice of SM radiation. The final slice is the uniform-density slice of GWs and neutrinos.
Then, the relation 
\begin{equation}
    \rho_{\text{GW},i}(\vb{x}) e^{-4N(\vb{x})} + \rho_{\nu,i} e^{-4N(\vb{x})} = \rho_{\text{GW} + \nu, f}
\end{equation}
implies 
\begin{equation}
    \zeta_\nu = \delta N = \frac{1}{4} \frac{\delta \rho_\text{GW}}{\rho_{\nu} + \rho_{\text{GW}}},
\end{equation}
where $\delta \rho_\text{GW}$ is defined in the flat slice, or equivalently, in the uniform-density slice of SM radiation.
Since the GW density typically has a power-law dependence $\rho_\text{GW} \propto k_\text{re}^{\alpha}$ and $\delta k_\text{re}$ is determined by $\sigma$ fluctuation as shown in \cref{eqn:sigmaOfK}, one can relate
\begin{equation}
    \rho_\text{GW} = \bar{\rho}_\text{GW} \qty(\frac{\sigma(k)}{\bar{\sigma}(k)})^{\alpha/\beta}.
    \label{eqn:rhoGWfluctuation}
\end{equation}
Hence, to quadratic order in the Gaussian fluctuation $\delta \sigma$, the GW curvature perturbation is
\begin{equation}
    \begin{aligned}
        \zeta_\nu =& \fdv{N}{\rho_\text{GW}} \fdv{\rho_\text{GW}}{\sigma} \delta \sigma + \frac{1}{2} \left[ \fdv[2]{N}{\rho_\text{GW}} \qty(\fdv{\rho_\text{GW}}{\sigma})^2 \right. \\
        & \left. + \fdv{N}{\rho_\text{GW}} \fdv[2]{\rho_\text{GW}}{\sigma} \right] \delta \sigma \delta \sigma \\
        \approx& \frac{1}{4} \frac{\Omega_\text{GW}}{\Omega_\nu} \frac{\alpha}{\beta} \frac{\delta \sigma}{\sigma} + \frac{1}{8} \frac{\Omega_\text{GW}}{\Omega_\nu} \qty(\frac{\alpha}{\beta})^2 \qty(\frac{\delta\sigma}{\sigma})^2,
    \end{aligned}
\end{equation}
in which we used the following relations 
\begin{equation}
    \begin{gathered}
        \fdv{N}{\rho_\text{GW}} = \frac{1}{4(\rho_\text{GW} + \rho_\nu)} \approx \frac{1}{4\rho_\nu}, \quad 
        \fdv[2]{N}{\rho_\text{GW}} \approx -\frac{1}{4\rho_\nu^2}, \\ 
        \fdv{\rho_\text{GW}}{\sigma} = \frac{\alpha}{\beta} \frac{\rho_\text{GW}}{\sigma}, \quad 
        \fdv[2]{\rho_\text{GW}}{\sigma} \approx \frac{\alpha^2}{\beta^2} \frac{\rho_\text{GW}}{\sigma^2},
    \end{gathered}
\end{equation}
and assumed that $\rho_\text{GW} \ll \rho_\nu$ to be compatible with the $\Delta N_\text{eff}$ constraint from \textit{Planck} \cite{Planck:2018vyg} and $\beta \ll 1$ so that $\sigma$ undergoes a slow roll. Therefore, the effective NDI perturbation due to GW from $\sigma$-modulated re-entry of cosmic strings is 
\begin{align}
    \Delta_\text{NDI}^2 =& \qty(\frac{3\Omega_\text{GW}}{4\Omega_\nu} \frac{\alpha}{\beta})^2 \Delta_\sigma^2 
    \approx \qty(\frac{3 \Omega_\text{rad}}{4\Omega_{\nu}})^2 \frac{\delta \Omega^2_\text{GW}(f_\text{peak})}{\Omega_\text{rad}^2} \nonumber \\
    \approx& \qty( 1.9 \cdot \frac{\delta \Omega_\text{GW}(f_\text{peak})}{\Omega_\text{rad}} )^2.
    \label{appeqn:stringNDItoGW}
\end{align}
where $f_\text{peak}$ is the frequency at which the GW anisotropy is maximized. Here, we used \cref{eqn:rhoGWfluctuation} to relate $\Delta_\sigma$ to $\delta\Omega_\text{GW}$. Using the \textit{Planck} constraint at $k_\text{low} = \SI{0.002}{\Mpc^{-1}}$ in which $\Delta^2_\text{NDI}(k_\text{low}) \lesssim 7.4\%\Delta_\zeta^2 \lesssim 1.6\times 10^{-10}$, we find that 
\begin{equation}
    \delta \Omega_\text{GW}(f_\text{peak}) h^2 \lesssim 0.15 \cdot \Omega_\text{rad} h^2 \Delta_\zeta(k_\text{low}) \approx 2.8\times 10^{-10}.
\end{equation}
However, if we take the NANOGrav constraint on stable cosmic strings $G \mu \lesssim 10^{-9.5}$ \cite{NANOGrav:2023hvm}, then we conclude that $\Omega_\text{GW}/\Omega_\text{rad} \lesssim 6.4\times 10^{-7}$. Hence, if the NDI constraint is saturated, the GW fluctuation is at least
\begin{equation}
    \frac{\delta\Omega_\text{GW}(f_\text{peak})}{\Omega_\text{GW}} \approx 0.15 \qty(\frac{\Omega_\text{rad}}{\Omega_\text{GW}}) \Delta_\zeta \gtrsim 11,
\end{equation}
which violates perturbativity. Therefore, requiring $\delta\Omega_\text{GW}(f_\text{peak}) / \Omega_\text{GW} <1$ for the cosmic-string-generated GW anisotropy and the NANOGrav constraints are typically sufficient to evade the current isocurvature constraint.

\subsubsection{Isocurvature Non-Gaussianity}
Beyond two-point correlation, it is also possible that an NDI non-Gaussianity is produced. The isocurvature fluctuation $S_\text{NDI}(k)$ is related to the underlying almost Gaussian field $\delta_\sigma = \delta\sigma/\sigma$ by 
\begin{equation}
    S_\text{NDI}(\vb{k}) = \frac{3}{4} \frac{\Omega_\text{GW}}{\Omega_\nu} \frac{\alpha}{\beta} \delta_\sigma(\vb{k}) + \frac{3}{8} \frac{\Omega_\text{GW}}{\Omega_\nu} \qty(\frac{\alpha}{\beta})^2 (\delta_\sigma \star \delta_\sigma)(\vb{k}).
\end{equation}
in which $\star$ denotes a momentum convolution. Then, the tree-level isocurvature bispectrum is obtained by computing $\ev{S_\text{NDI}^3}$ of three different momenta
\begin{align}
    & B_{iii}(\vb{k}_1, \vb{k}_2, \vb{k}_3) \\
    =& \frac{3^3}{4^2 \cdot 8} \qty(\frac{\Omega_\text{GW}}{\Omega_\nu})^3 \qty(\frac{\alpha}{\beta})^4 \qty[ 2 \Delta_\sigma^2(k_1) \Delta_\sigma^2(k_2) + \text{2 perms} ] \nonumber \\
    =& f_\text{NL}^{i,ii} \qty[ 2 \Delta_\Phi^2(k_1) \Delta_\Phi^2(k_2) + \text{2 perms} ]. \nonumber 
\end{align}
For the cosmic string scenario, the gravitational wave spectrum is related to the NDI spectrum by \cref{appeqn:stringNDItoGW} so that the local-type $f_\text{NL}$ is
\begin{align}
    f_\text{NL}^{i,ii} = & \frac{625}{384} \qty(\frac{\Omega_\text{GW}}{\Omega_\nu})^3 \qty(\frac{\delta \Omega_\text{GW}(f_\text{peak})}{\Omega_\text{GW} \Delta_\zeta})^4 \\
    \approx& 1.4 \cdot \qty(\frac{\Omega_\text{GW}}{6.4\times 10^{-7} \Omega_\text{rad}})^3 \qty(\frac{\delta\Omega_\text{GW}/\Omega_\text{GW}}{1})^4, \nonumber
\end{align}
where we relates the curvature perturbation $\zeta $ to the gravitational potential $\Phi = 3 \zeta /5$.
The current constraint on the NDI non-Gaussianity from \textit{Planck} is $\abs{f_\text{NL}^{i,ii}} \lesssim 210$~\cite{Planck:2019kim}.
As long as the NANOGrav constraint $\Omega_\text{GW}/\Omega_\text{rad} \lesssim 6.4\times 10^{-7}$ is satisfied, the non-Gaussianity constraint is also satisfied.

It is possible that cosmic strings disappear by the nucleation or reentry of monopoles~\cite{Martin:1996ea, Martin:1996cp} by the time the NANOGrav-scale GWs would be emitted. In this case, $G\mu$ may be larger and the NDI and/or non-Gaussianity may be large enough to be observed in future observations.

To our knowledge, this setup produces the largest GW anisotropies among the models known in the literature, without producing excessive cosmic perturbations or their non-Gaussianity. This is because almost all the energy of the source of GWs, namely, cosmic strings, goes to GWs rather than other forms of radiation \cite{Moore:1998gp, Olum:1999sg}.%
\footnote{This argument comes from the standard Nambu-Goto approximation of cosmic strings \cite{Vilenkin:2000jqa} and is supported by Abelian-Higgs simulations \cite{Moore:1998gp, Olum:1999sg}. However, it is worth noting that there is debate over whether it is possible to primarily produce small string loops that can radiate massive modes \cite{Antunes:1997pm, Hindmarsh:2008dw}. This leads to the small-string-loop scenario. More details about the large-string-loop and small-string-loop scenario are discussed in Ref.~\cite{Auclair:2019jip}. As our discussion focuses on semi-classical large-scale dynamics, we thus argue that gravitational wave radiation is the dominant decay channel for consistency. }

\section{Gravitational-Wave Anisotropy Signal from Domain Walls}\label{sec:GWfromWall}

\subsection{GW Signal}
\label{subsec:WallSignal}
In this section, we consider another possibility that domain walls form during inflation and are pushed outside the horizon. For simplicity, we will assume that upon their reentry, the wall network immediately starts to annihilate due to bias between the false and true vacua. For concrete realization, we consider the following Lagrangian
\begin{equation}
    -\Lag \supset \frac{1}{2}\qty( \lambda_{\sigma\chi}^2 \sigma^2 - m_\chi^2) \chi^2 + \frac{\lambda_\chi^2}{4} \chi^4 + \varepsilon^3 \chi
\end{equation}
with $\varepsilon \ll m_\chi$. The potential bias across the two vacua $ \delta \sim 2\mVac \varepsilon^3/\lambda_\chi$, while the wall tension $\tau \sim  \mVac^3 /\lambda_\chi^2 $. The walls immediately collapse after their horizon reentry if $ \tau H_\text{re} < \delta$, which requires $H_\text{re} < \lambda_\chi \varepsilon (\varepsilon/\mVac)^2 $.
As $H_\text{re}$ is typically small due to the inflationary dilution, it is easy to realize this inequality.

The GW spectrum can be estimated as 
\begin{equation}
    \begin{multlined}
        \Omega_\text{GW} = \Omega_\text{rad} \frac{0.7 \cdot \tau^2 k_\text{eq}^4}{24\pi \MPl^4 H_\text{eq}^2 k_\text{re}^4} 
        \times
        \begin{dcases}
            \qty(\frac{2\pi f}{k_\text{re}})^{3}, & f \leq \frac{k_\text{re}}{2\pi}, \\
            \qty(\frac{k_\text{re}}{2\pi f})^{n_\text{UV}}, & f > \frac{k_\text{re}}{2\pi}, \\
        \end{dcases}
    \end{multlined}
\end{equation}
in which we used $H_\text{re} = H_\text{eq} (k_\text{re}/k_\text{eq})^2$ during RD. The spectral index $n_\text{UV}$ is determined by detailed UV physics and is extracted numerically, and $n_\text{UV} \approx 1$ for scaling walls that collapse due to bias \cite{Hiramatsu:2013qaa}. Our scenario slightly differs because the bias is already large so that the wall network collapses immediately upon reentry. Nonetheless, we expect a similar, if not sharper, UV rolloff. Then, the anisotropy signal is
\begin{align}
    &\OmegaGWaniso(\ell, f) \sim \Omega_\text{rad} \frac{0.7 \cdot \tau^2 k_\text{eq}^4}{24 \MPl^4 H_\text{eq}^2 k_\text{re}^4} \qty(\frac{\lambda_{\sigma\chi} H_I}{2\pi \mInf}) \\
        &\times \frac{1}{\beta} \qty(\frac{\ell}{k_\text{re} D_*})^\beta \times
        \begin{dcases}
            7 \cdot \qty(\frac{2\pi f}{k_\text{re}})^3, & f \leq \frac{k_\text{re}}{2\pi}, \\
            \qty(4 - n_\text{UV}) \qty(\frac{k_\text{re}}{2\pi f})^{n_\text{UV}}, & f > \frac{k_\text{re}}{2\pi}. 
        \end{dcases} \nonumber
\end{align}

\subsection{Constraints}
\label{subsec:WallConstraints}
Similar to the cosmic-string case, large cosmic perturbations can be generated by the modulation of domain walls' reentry. As discussed below, we obtain a constraint 
\begin{align}
    & \delta \Omega_\text{GW}(f_\text{peak}) h^2 \lesssim \min\left\lbrace 2.8, 4.0 \sqrt{\frac{\Omega_\text{GW}}{2.3\% \Omega_\text{rad}}}, \right. \\
    & \left. 4.3 \qty(\frac{\Omega_\text{GW}}{2.3\% \Omega_\text{rad}})^{5/8},  0.99\times \qty(\frac{\Omega_\text{GW}}{2.3\%\Omega_\text{rad}})^{1/4} \right\rbrace \times10^{-10}. \nonumber
\end{align}
The first constraint comes from the bound on GW isocurvature perturbations similar to the cosmic-string scenario, the second is due to the adiabatic perturbations from the radiation created by domain walls' collapse, the third is from their non-Gaussianity, and the fourth is the GW isocurvature perturbations partially correlated with the curvature perturbations. This constraint is more stringent than the string case because domain walls dominantly decay into non-GW components, which obtain larger fluctuations than GWs.%
\footnote{The case (ii) of the curvaton model in~Ref.~\cite{Malhotra:2022ply} will be subject to the analogous constraints. If the first constraint is the strongest, the setup in~Ref.~\cite{Malhotra:2022ply} can generate as large $\delta \Omega_\text{GW}$ as our string case when the $\Delta N_\text{eff}$ constraint is nearly saturated.}

\subsubsection{Adiabatic Power Spectrum} 

For the domain wall (DW) that collapses immediately after reentry, the wall predominantly decays into other Standard Model radiation species rather than GWs. This is because the gravitational wave emission rate is $\sim \tau/\MPl^2$ in which $\tau$ denotes the wall tension while the decay rate of the network is $\sim H_\text{re}$ (immediate collapse upon reentry) so that $\rho_\text{GW} \approx \epsilon \tau \rho_\text{DW}/(8\pi \MPl^2 H_\text{re})$ around walls' reentry $t_\text{re}$ in which the efficiency factor $\epsilon \approx 0.7$ \cite{Hiramatsu:2013qaa} and $\rho_\text{DW}(t_\text{re}) = \tau H_\text{re}$. If DWs' reentry happens during RD, we can express 
\begin{equation}
    \begin{multlined}
        \frac{\rho_\text{GW}}{\rho_\text{rad}} = \frac{3\epsilon}{8\pi} \qty(\frac{\rho_\text{DW}}{\rho_\text{rad}})^2 \\
        \implies 
        \frac{\rho_\text{DW}}{\rho_\text{rad}} = \qty(\frac{g_{*,\text{re}}}{g_{*,0}})^{1/6} \sqrt{\frac{8\pi}{3\epsilon} \frac{\Omega_\text{GW}}{\Omega_\text{rad}}},
    \end{multlined}
    \label{appeqn:DWradRatio}
\end{equation}
where $\rho_\text{DW}/\rho_\text{rad}$ is evaluated around the reentry while $\Omega_\text{GW}/\Omega_\text{rad}$ is evaluated today.

On the other hand, the SM radiation from domain walls observed at a later time $t_f$ will obtain an energy density $\rho_\text{DW}(t_f) = \tau H_\text{re} (T_f/T_\text{re})^4 \propto T_\text{re}^{-2}$. Thus, we can track the curvature perturbation by treating the original SM radiation and that generated by DW decay separately as a radiation fluid with density $\rho_\text{SM}$ and another with $\rho_\text{DW}$. Assuming $\rho_\text{DW} \ll \rho_\text{SM} \approx \rho_\text{rad}$ (which is required to avoid domain-wall-dominated epoch), we may take the initial flat slice at some temperature $T_* < T_\text{re}$ of the SM radiation and take the final slice at a uniform-density slice of SM + DW fluid and enforce
\begin{equation}
    T_*^4 e^{-4N(\vb{x})} + \frac{\tau}{\MPl} T_\text{re}^2(\vb{x}) \qty(\frac{T_*}{T_\text{re}(\vb{x})})^4 e^{-4N(\vb{x})} = \rho_\text{rad},
\end{equation}
which implies that the curvature perturbation from DW-generated radiation is 
\begin{equation}
    \zeta_\text{DW} = -\frac{1}{2} \frac{\rho_\text{DW}}{\rho_\text{rad}} \frac{\delta T_\text{re}}{T_\text{re}} = \frac{1}{4} \frac{\delta \rho_\text{DW}}{\rho_\text{rad}}.
\end{equation}
in which we used the relation that $\delta T_\text{re}/T_\text{re} = -\delta \rho_\text{DW}/(2 \rho_\text{DW})$. And a subdominant curvature perturbation due to GW is 
\begin{equation}
    \begin{multlined}
        \frac{\delta \rho_\text{GW}}{\rho_\text{GW}} 
        = 2\frac{\rho_\text{rad}}{\rho_\text{DW}} \frac{\delta \rho_\text{DW}}{\rho_\text{rad}}
        = 8 \frac{\rho_\text{rad}}{\rho_\text{DW}} \zeta_\text{DW} \\
        = \qty(\frac{g_{*,0}}{g_{*,\text{re}}})^{1/6} \sqrt{\frac{24\epsilon}{\pi} \frac{\Omega_\text{rad}}{\Omega_\text{GW}}} \zeta_\text{DW},
    \end{multlined}
\end{equation}
in which we used \cref{appeqn:DWradRatio}. Thus, the adiabatic constraint on $\zeta_\text{DW}$ limits the anisotropic GW signal to be below
\begin{align}
    & \delta \Omega_\text{GW}(f_\text{peak}) h^2 \lesssim \qty(\frac{g_{*,0}}{g_{*, \text{re}}})^{1/6} \sqrt{\frac{24\epsilon}{\pi} \frac{\Omega_\text{GW}}{\Omega_\text{rad}}} \Omega_\text{rad} h^2 \Delta_\zeta(k_\text{low}) \nonumber \\
    \approx& 4.0\times 10^{-10} \sqrt{\frac{\Omega_\text{GW}}{2.3\% \Omega_\text{rad}}} \qty(\frac{106.75}{g_{*,\text{re}}})^{1/6}.
\label{eq:DW_ad}
\end{align}
When the $\Delta N_{\rm eff}$ constraint is saturated by taking $\Omega_\text{GW} = 2.3\%\Omega_\text{rad}$, this is a weaker constraint than the NDI constraint $\delta \Omega_\text{GW}(f_\text{peak}) h^2 \lesssim 2.8\times 10^{-10}$. However, for generically lower isotropic background $\Omega_\text{GW}$, this is a stronger constraint. Alternatively, one may compare the constraint on the GW fluctuation, \textit{i.e.},
\begin{equation}
    \frac{\delta \Omega_\text{GW}(f_\text{peak})}{\Omega_\text{GW}} \lesssim 4.2\times 10^{-4} \sqrt{\frac{2.3\% \Omega_\text{rad}}{\Omega_\text{GW}}} \qty(\frac{106.75}{g_{*,\text{re}}})^{1/6}.
\end{equation}
The adiabatic-perturbation constraint remains strong until $\Omega_\text{GW} h^2 \lesssim 1.7 \times 10^{-13}$ at which the perturbativity constraint $\delta \Omega_\text{GW} < \Omega_\text{GW}$ becomes effective.

\subsubsection{Adiabatic Non-Gaussianity} 
Non-Gaussianity of the DW-originated fluctuations can be obtained by expanding to 2\textsuperscript{nd}-order in $\delta \sigma$,
\begin{equation}
    \zeta_\text{DW} \approx \frac{1}{2\beta} \frac{\rho_\text{DW}}{\rho_\text{rad}} \frac{\delta \sigma}{\sigma} + \frac{1}{2\beta^2} \frac{\rho_\text{DW}}{\rho_\text{rad}} \frac{\delta \sigma^2}{\sigma^2},
\end{equation}
leading to a bispectrum of the form
\begin{align}
    & B_{aaa}(\vb{k}_1, \vb{k}_2, \vb{k}_3) \\
    =& \frac{1}{2^3 \beta^4} \qty(\frac{\rho_\text{DW}}{\rho_\text{rad}})^3 \qty[ 2\Delta^2_\sigma(k_1)\Delta^2_\sigma(k_2) + \text{2 perms} ]. \nonumber \\
    =& f_\text{NL}^{a,aa} \qty[ 2 \Delta_\Phi^2(k_1) \Delta_\Phi^2(k_2) + \text{2 perms} ]. \nonumber 
\end{align}
Its corresponding local $f_\text{NL}$ is
\begin{equation}
    \begin{aligned}
        f_\text{NL}^{a,aa} \approx& \frac{5}{3} \frac{1}{2^3 \beta^4} \qty(\frac{\rho_\text{DW}}{\rho_\text{rad}})^3 \qty(\frac{\beta}{4})^4 \qty(\frac{\delta\Omega_\text{GW}(f_\text{peak})}{\Omega_\text{GW} \Delta_\zeta})^4 \\
        =& \frac{5}{12} \sqrt{\frac{g_{*,\text{re}}}{g_{*,0}}} \qty(\frac{\pi}{24 \epsilon} \frac{\Omega_\text{GW}}{\Omega_\text{rad}})^{3/2} \qty(\frac{\delta \Omega_\text{GW}(f_\text{peak})}{\Omega_\text{GW} \Delta_\zeta})^{4}
    \end{aligned}
\end{equation}
assuming $\Delta_\text{DW}^2 \ll \Delta_\zeta^2$ which holds when $\qty(\rho_\text{DW}/\rho_\text{rad}) \qty(\dot{\phi}/\dot{\sigma}) \ll 1$. Recasting the \textit{Planck} constraint  $\abs{f_\text{NL}^{a,aa}} \lesssim 5.1$ \cite{Planck:2019kim}, we find that the anistropic GW signal must be smaller than
\begin{equation}
    \delta\Omega_\text{GW}(f_\text{peak}) h^2 \lesssim 4.3\times10^{-10} \qty(\frac{\Omega_\text{GW}}{2.3\%\Omega_\text{rad}})^{5/8} \qty(\frac{106.75}{g_{*,\text{re}}})^{1/8}.
\end{equation}

\subsubsection{Correlated Isocurvature} 
Lastly, this type of signal also induces correlated isocurvature. The correlation is
\begin{align}
     & \ev{S_\text{NDI} \zeta_\text{DW}} 
     = \frac{3}{4} \qty(\frac{g_{*,\text{re}}}{g_{*,0}})^{1/6} \sqrt{\frac{\pi}{24\epsilon}} \frac{\Omega_\text{rad}}{\Omega_\nu} \nonumber \\
     & \times \qty(\frac{\Omega_\text{GW}}{\Omega_\text{rad}})^{3/2} \qty(\frac{\delta\Omega_\text{GW}(f_\text{peak})}{\Omega_\text{GW}})^2.
\end{align}
The \textit{Planck} constraint $\ev{S_\nu \zeta_\text{DW}} < 5\times 10^{-11}$~\cite{Planck:2018jri} requires
\begin{align}
    \delta \Omega_{\text{GW}} h^2 \lesssim 9.9 \times10^{-11} \qty(\frac{\Omega_\text{GW}}{2.3\%\Omega_\text{rad}})^{1/4} \qty(\frac{106.75}{g_{*,\text{re}}})^{1/12}.
\end{align}
Although the correlated isocurvature constraint is the strongest one for the benchmark of $\Omega_\text{GW}$ here, because the $f_\text{NL}^{a,aa}$ constraint $\propto \Omega_\text{GW}^{5/8}$ drops quicker than $\Omega_\text{GW}^{1/4}$ as $\Omega_\text{GW}$ decreases, the $f_\text{NL}^{a,aa}$ constraint eventually becomes stronger at small $\Omega_\text{GW}$. 

\section{Symmetry Breaking by Two Fields}
\label{app:two fields}

So far, we consider a case where a charged field breaks symmetry during inflation and remains the dominant source of symmetry breaking also in the vacuum. This setup may be slightly varied so that the symmetry-breaking field during inflation is different from that in the vacuum, which is realized in an embedding to the minimal supersymmetric SM (MSSM) discussed in \cref{app:SUSY}. 

\begin{figure}
    \centering
    \includegraphics[width=0.4\textwidth]{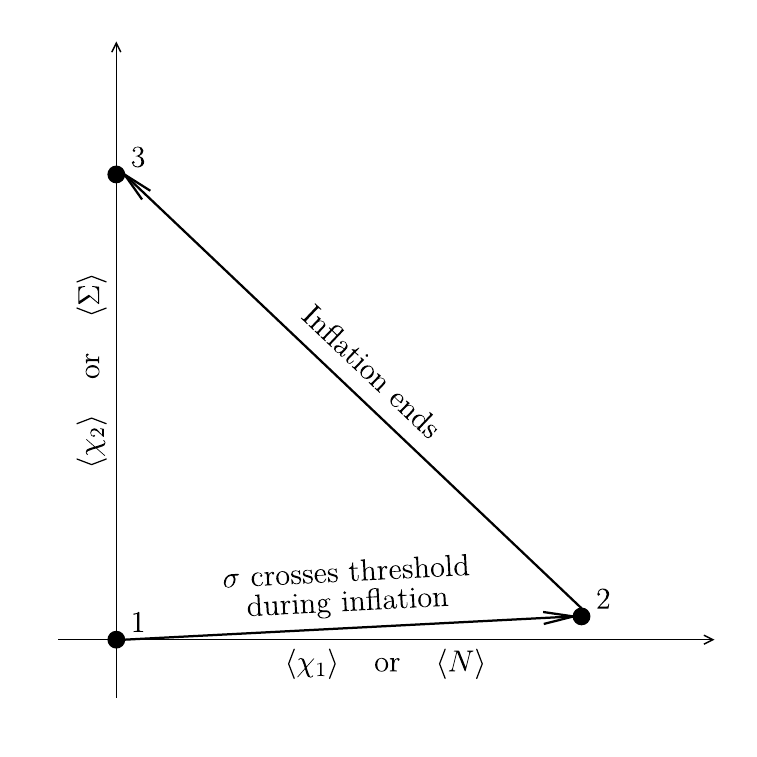}
    \caption{Schematic for the evolution of the field values in the two-field symmetry breaking discussed in \cref{app:two fields}. Initially, both charge-1 field $\chi_1$ and charge-2 field $\chi_2$ have no field values during inflation. When the spectator field $\sigma$ crosses a threshold, $\chi_1$ rolls to a minimum generated by the tachyonic Hubble-induced mass. $\chi_2$ field may obtain a tiny field value by coupling to $\chi_1$. After inflation, $\chi_1$'s Hubble-induced mass term becomes negligible while $\chi_2$'s tachyonic vacuum mass becomes significant and drives $\chi_2$ to a new field value. In the supersymmetric realization, $\chi_1$ can be identified with right-handed sneutrino $N$, $\chi_2$ with the $B-L$ breaking scalar, and $\sigma$ with the Higgs field as discussed in \cref{app:SUSY}.}
    \label{fig:VEVevol}
\end{figure}

For example, let us introduce two fields $\chi_1$ and $\chi_2$, where the subscript shows the $\U1$ charge of each field, and consider the Lagrangian of the form 
\begin{align}
    -\Lag = &\qty(\lambda_{\sigma \chi}^2 \sigma^2 - c_1 H_I^2 + m_1^2) \abs{\chi_1}^2 + \frac{\lambda_1^2}{4} \abs{\chi}^4 \\ + &\qty(c_2 H_I^2 - m_2^2) \abs{\chi_2}^2 + \frac{\lambda_2^2}{4}\abs{\chi_2}^4 + \qty( \Lambda \chi_1^2 \chi_2^* + \hc). \nonumber
\end{align}
During inflation, assuming $c_1 H_I^2 > m_1^2$, $\chi_1$ breaks the $\U{1}$ symmetry and produces cosmic strings. Assuming $c_2 H_I^2 > m_2^2$, $\chi_2$ is trapped near the origin, but
because of the interaction term between $\chi_1$ and $\chi_2$, $\chi_2$ obtains a non-zero field value.
Parameterizing the fields as $\chi_i \sim v_i e^{\ii \theta_i}$, the interaction term $\propto \cos(\theta_2 - 2\theta_1)$, so $\chi_2$ winds twice more around the cosmic strings than $\chi_1$, which $\chi_1$ typically winds once around strings. Then, as the Hubble-induced mass of $\chi_1$ becomes small after inflation, the VEV of $\chi_1$ vanishes while $\U{1}$ continues to be broken by $\chi_2$, with cosmic strings made by $\chi_2$. The evolution of the field values is sketched in \cref{fig:VEVevol}.

Two-field breaking can also be realized in the domain wall scenario. In this case, $\chi_1$ and $\chi_2$ are real fields and charged under some $\Z[2]$. With the interaction term replaced by $\Lambda^2 \chi_1 \chi_2$, one can alter the wall-forming field during inflation from that in the vacuum as well.

\section{Model Realization: GW Anistropy from Higgs-driven $B-L$ Strings}
\label{app:SUSY}

So far, we have discussed our mechanism within toy models. We now embed the mechanism into a well-motivated UV extension of the SM. To this end, we consider supersymmetric theory, in which the lightness of $\sigma$ and $\chi$ can be understood in terms of supersymmetry.

To be concrete, we consider $\U{1}[$B-L$]$ phase transition and shows that the Higgs fields in the MSSM can play the role of $\sigma$. Besides the MSSM fields, we introduce a right-handed neutrino field $N$ with $B-L$ charge $1$, a scalar $\Sigma$ with $B-L$ charge $-2$, and another $\bar{\Sigma}$ field with $B-L$ charge $2$ to ensure anomaly cancellation. These fields will follow a two-field symmetry breaking dynamics, a simplified version of which is discussed in \cref{app:two fields}. We start by considering the following superpotential
\begin{equation}
    W = y L H_u N + \frac{\lambda}{2} \Sigma N^2.
\end{equation}
This leads to the following Lagrangian
\begin{equation}
    \begin{aligned}
        - \Lag \supset& \qty(\abs{y}^2 \abs{H_u}^2 - c_N H_I^2 + \msoft^2) \abs{N}^2 \\
        & + \qty(\abs{\lambda}^2 \abs{N}^2 + c_\Sigma H_I^2 - \msoft^2) \abs{\Sigma}^2 \\
        & + \qty(c_{\bar{\Sigma}} H_I^2 - \msoft^2) \abs{\bar{\Sigma}}^2 \\
        & + \qty(\lambda \msoft \Sigma N^2 + \hc) \\ 
        & + \frac{\abs{\lambda}^2}{4} \abs{N}^4 + \frac{g_{B-L}^2}{2} \qty(\abs{N}^2 - 2 \abs{\Sigma}^2 + 2 \abs{\bar{\Sigma}}^2)^2,
    \end{aligned}
\end{equation}
where we took $L=0$, which can be guaranteed by a large enough positive mass of $L$. We omit the $D$-term potential by the MSSM gauge group, and in the following, $H_u$ is understood as a $D$-flat direction $H_u H_d$ that plays the role of the light spectator. Heavy neutral slepton $N$ is the field $\chi_1$ undergoing phase transition during inflation, and $\Sigma$ is the field $\chi_2$ that breaks $\U{1}[$B-L$]$ at the vacuum.

Initially, $H_u$ has a large field value but is light during inflation, while other fields have vanishing field values. As it scans to the critical value $H_u \lesssim \sqrt{c_N} H_I/y$, $N$ undergoes a phase transition, producing $B-L$ gauge strings. We may assume that $\lambda < g_{B-L}$ such that the $D$-term potential of $N$ field dominates and fixes $N$ to obtain a field value $\ev{N} \simeq \sqrt{c_N} H_I/g_{B-L}$. By setting $g_{B-L} < 1$ or $c_N > 1$, $N$ can be fixed to a field value larger than $H_I$ with a larger mass so that its de Sitter fluctuation can be suppressed, and $N$'s phase transition is complete during inflation. In the meantime, the $\Sigma$ field is also nudged away from zero because of the $A$-term potential. The new field value of $\Sigma$ is $\ev{\Sigma} \simeq c_N \lambda \msoft/(c_\Sigma g_{B-L}^2)$ with a winding dictated by the winding of $N$ field. Hence, $\Sigma$ strings are formed during inflation as well. $\bar{\Sigma}$ does not need to obtain a field value yet, which can be accommodated by a positive Hubble-induced mass. 

After inflation, $\ev{N}$ decreases due to the positive vacuum mass. $\ev{\Sigma}$ and $\ev{\bar{\Sigma}}$ increase because of negative soft mass terms. $D$-flatness ensures that $\ev{\Sigma} \sim \ev{\bar{\Sigma}}$. To stabilize the $\ev{\Sigma}$ and $\ev{\bar{\Sigma}}$, we have two options: 1) introducing another singlet $X$ or 2) considering a radiative breaking of $\U{1}[$B-L$]$. 

The following superpotential realizes the first case
\begin{equation}
    W \supset \kappa X(\Sigma \bar{\Sigma} - v^2)
\end{equation}
so that the $F$ term of $X$ forms a wine-bottle potential $\sim \kappa^2 \abs{\Sigma \bar{\Sigma} - v^2}^2$. If $\kappa \lesssim \sqrt{c_\Sigma + c_{\bar{\Sigma}}} H_I/v$, the tachyonic mass term is negligible during inflation and will not affect the symmetry-breaking pattern discussed in the previous paragraphs. Once $\ev{\Sigma} \sim \ev{\bar{\Sigma}} \sim v$, $N$ strings formed during inflation will be wrapped around by $\Sigma$ and $\bar{\Sigma}$ field, and the $D$-term potential can ensure a safe transfer of the winding. Although the $A$-term potential winds the $\Sigma$ field twice around the $N$ field, the non-minimal $\Sigma$ strings remain stable until reentry because $g_{B-L} > \kappa$ leads to an attractive force binding two minimal $\Sigma$ strings together. Also, the post-inflationary breaking of $\bar{\Sigma}$ can produce another cosmic string due to the breaking of the global chiral symmetry of $\Sigma$ and $\bar{\Sigma}$. This string, however, is not stable because the $\kappa^2 v^2 \Sigma \bar{\Sigma}$ term explicitly breaks the chiral symmetry. 

More minimally, one may assume that the soft mass of $\Sigma$ gets radiatively corrected to obtain a minimum~\cite{Moxhay:1984am}
\begin{equation}
    -\Lag \simeq \msoft^2 \qty[ 1 + \qty(\frac{\lambda}{4\pi})^2 \ln(\frac{\Sigma^2}{\Lambda^2}) ] \Sigma^2, 
\end{equation}
which yields $\ev{\Sigma} \simeq \Lambda \exp(-\qty(4\pi/\lambda)^2)$. In this case, the breaking of $\bar{\Sigma}$ string produces a stable global string besides the $\U{1}[$B-L$]$ string, and the Nambu-Goldstone boson of this global symmetry can be massless without further explicit breaking. However, as long as $v \lesssim 10^{-2}\MPl$, the Nambu-Goldstone radiation from global strings can evade the $\Delta N_\text{eff}$ constraint as the string network in the scaling regime is a subdominant component of the radiation bath.

We considered the case where $H_u H_d$ is $\sigma$ and $N$ is $\chi$, but other combinations are possible. For example, with $H_u H_d$ being $\sigma$,  $LL\bar{e}$ may be $\chi$.
Also, $X$ may play the role of $\sigma$ with $\Sigma \bar{\Sigma}$ being $\chi$.

\section{Conclusion}
\label{sec:conclusion}
In this work, we have demonstrated that a phase transition modulated by a spectator field during inflation can generate sizable anisotropies in the GW background. As an explicit realization, we considered a scenario in which the timing of a phase transition during inflation is modulated by spectator-field fluctuations, leading to the formation of topological defects whose subsequent horizon reentry produces anisotropic GW signals. Although the associated (iso)curvature perturbations impose constraints on the amplitude of the anisotropy, our framework allows for $\sim\mathcal{O}(1)$ modulation of GWs on large scales. This mechanism can be naturally embedded in supersymmetric theories, where the spectator field may be identified with the Higgs field of the MSSM.

The spectator-modulated phase transition offers several advantages over scenarios in which the phase transition is modulated by the inflaton. In inflaton-modulated models, both the adiabatic perturbations and the GW anisotropies originate from inflaton fluctuations, making it difficult to generate large GW anisotropies while maintaining the observed small amplitude of adiabatic perturbations. Moreover, inflaton-modulated phase transitions typically require a large excursion of the inflaton field in order to induce a significant variation in the mass of the symmetry-breaking field during inflation. These requirements are absent in the spectator-based framework.

While our analysis focused on phase transitions triggered by the slow-roll dynamics of a spectator field, other triggering mechanisms are possible. For example, Ref.~\cite{Garbrecht:2025crs} studied quantum tunneling triggered by the decreasing expansion rate during inflation, where the bounce action is modulated by spectator-field fluctuations to generate curvature perturbations. If such quantum tunneling induces a phase transition, it may lead to GW signals qualitatively similar to those discussed here.

More broadly, this framework readily interfaces with a wide range of particle-physics and cosmological mechanisms. Phase transitions associated with grand unified theories, the Peccei–Quinn mechanism, or the Affleck–Dine mechanism may occur during inflation rather than before it. The framework may also be combined with other sources of GWs, such as scalar-induced tensor perturbations, to explore richer phenomenology. Finally, a spectator-modulated phase transition may itself account for the observed curvature perturbations, realizing a new class of curvaton-like models. All of these possibilities suggest that spectator-modulated phase transitions open a promising avenue for exploring the interplay between inflationary dynamics, novel cosmological observables, and fundamental particle-physics questions.

\acknowledgments
We thank the enlightening comments from Soubhik Kumar. YB and KH are supported by the Department of Energy grant DE-SC0009924. KH is also supported by World Premier International Research Center Initiative (WPI), MEXT, Japan (Kavli IPMU).

\appendix

\section{Sensitivities for GW Anisotropy}
\label{app:GWSensitivity}
In this appendix, we detail how sensitivities of current and future GW observatories are recasted in this paper. We match our $\OmegaGWaniso(\ell,f)$ to the spherical harmonic coefficients $\Omega_{\text{GW}, \ell}^m$ of the directional GW spectrum
\begin{equation}
    \Omega_\text{GW}(\hat{n}, f) = \sum_{\ell, m} \Omega_{\text{GW}, \ell}^m Y_{\ell}^m(\hat{n})
\end{equation}
on the celestial sphere \cite{LISACosmologyWorkingGroup:2022kbp, Cui:2023dlo}. Because $Y_0^0(\hat{n}) = 1/\sqrt{4\pi}$, the monopole coefficient is larger than the usual isotropic spectrum $\Omega_{\text{GW},\ell=0}^{m=0} = \sqrt{4\pi} \Omega_\text{GW}$. In other words, the sky-averaged spectrum $\Omega_\text{GW}(f) = (4\pi)^{-1} \int \dd^2 \hat{n} \; \Omega_\text{GW}(\hat{n}, f)$ is dimensionless while $\Omega_{\text{GW}, \ell}^m$ has a unit $\sr^{1/2}$ as $Y_{\ell}^m(\hat{n})$ has a unit $\sr^{-1/2}$. Another frequently used convention is to define $\Omega_\text{GW}(f) = \int \dd^2 \hat{n}\; \tilde{\Omega}_\text{GW}(\hat{n}, f)$ and perform an expansion using dimensionless spherical harmonic function $\tilde{Y}_{\ell}^m(\hat{n}) = \sqrt{4\pi} Y_{\ell}^m(\hat{n})$. In this expansion convention, the spherical harmonic coeffcient $\tilde{\Omega}_{\text{GW}, \ell}^m$ has a unit $\sr^{-1}$

Sometimes, one also introduces the relative directional fluctuation 
\begin{gather}
    \delta_\text{GW}(\hat{n}, f) = \frac{\Omega_\text{GW}(\hat{n},f) - \Omega_\text{GW}(f)}{\Omega_\text{GW}(f)}, \\ 
    \delta_{\text{GW}, \ell}^{m} = \int \dd^2 \hat{n} \; Y_{\ell}^{m*}(\hat{n}) \delta_\text{GW}(\hat{n}, f)
\end{gather}
in which $\delta_{\text{GW}, \ell}^{m}$ has a unit $\sr^{1/2}$ and define the variance of its spherical harmonic coefficient as 
\begin{equation}
    \ev{\delta_{\text{GW}, \ell}^{m*} \delta_{\text{GW}, \ell'}^{m'}} = \delta_{\ell' \ell} \delta_{m' m} C_{\ell}
\end{equation}
analogous to the CMB measurement so that 
\begin{equation}
    \Omega_{\text{GW}, \ell}^m \approx \sqrt{C_\ell} \Omega_\text{GW}(f)
\end{equation}
from a steradian counting. For the other convention widely used in GW observatories \cite{Thrane:2009fp, LIGOScientific:2025bkz}, one sometimes define a different $\tilde{C}_{\ell}$
\begin{equation}
    \ev{ \tilde{\Omega}_{\text{GW},\ell}^{m*} \tilde{\Omega}_{\text{GW},\ell'}^{m'} } = \delta_{\ell' \ell} \delta_{m' m} \tilde{C}_\ell,
\end{equation}
in which $\tilde{C}_{\ell}$ has a dimension $\sr^{-2}$. Then, the spherical harmonic coefficients relates to this $\tilde{C}_{\ell}$ via
\begin{equation}
    \Omega_{\text{GW}, \ell}^m \approx (4\pi)^{3/2} \sqrt{\tilde{C}_\ell}
\end{equation}
Lastly, much similar to the CMB $C_{\ell}$, the per-log-$\ell$ power spectrum is given by $\sqrt{\ell(\ell+1) C_{\ell}/(2\pi)}$ \cite{Cui:2019kkd, Bodas:2025wef}; therefore, when we plot the benchmark $\OmegaGWaniso(\ell,f)$ spectrum from modulated reentry of cosmic strings, we introduce another factor of $\sqrt{2\pi} [\ell(\ell+1)]^{-1/2}$ to convert the estimated per-log-$\ell$ $\OmegaGWaniso(\ell, f)$ to $\Omega_{\text{GW}, \ell}^m(f)$.

For sensitivity curves of GW observatories, it is common to project the reach using the power-law-integrated sensitivity (PLIS) curve $\Omega_\text{PLIS}$ \cite{Thrane:2013oya, Schmitz:2020syl, Bao:2025ori}. This reach is typically much deeper than per-frequency-bin (strain-noise) sensitivity as one usually integrate over multiple frequency bins to search a particular primordial signal. This is usually determined by the number of detectors $n_\text{det}$, observation time $t_\text{obs}$, effective noise spectrum $\Omega_\text{N}$, and a signal-to-noise-ratio threshold $\varrho_\text{thr}$. The PLIS is defined as 
\begin{equation}
    \begin{multlined}
        \Omega_\text{PLIS}(f) \\
        \defeq \max_{p}\left\lbrace \varrho_\text{thr} \left[ n_\text{det} t_\text{obs} \int_{f_\text{min}}^{f_\text{max}} \dd f\; \qty( \frac{(f/f_\text{ref})^p}{\Omega_\text{N}(f)} )^2 \right]^{-1/2} \right. \\
        \left. \times \qty(\frac{f}{f_\text{ref}})^p \right\rbrace.
    \end{multlined}
\end{equation}
Here, we pick $t_\text{obs} = 1\;\text{yr}$ and $\varrho_\text{thr} = 1$ to project reaches for future detectors, such as LISA, Einstein Telescope (ET), Big Bang Observer (BBO), and use the noise spectra from Refs.~\cite{LISACosmologyWorkingGroup:2022kbp, Cui:2023dlo} for anisotropic reach and Ref.~\cite{Schmitz:2020syl} for the isotropic reach. In addition, the first part of the fourth observing run of the LIGO-Virgo-KAGRA network (LVK O4a) also limits on the GW anisotropy on a scale-invariant spectrum \cite{LIGOScientific:2025bkz}. While different from PLIS, this still provides valuable insights on current exclusions. We cast these constraints on $\tilde{C}_\ell$ to $\Omega_{\text{GW}, \ell}^m$ and place them at the frequency ranges where the LVK detectors is the most sensitive. Lastly, we also used an ET sensitivity projection for the scale-invariant spectrum from Ref.~\cite{Mentasti:2020yyd} as a consistency check with other projections. 

\bibliography{ref_gw_anisotropy_from_light_spectator}

\end{document}